\documentclass[twocolumn, switch]{article} 

\usepackage{preprint}

\usepackage{amsmath, amsthm, amssymb, amsfonts}

\usepackage[numbers,square]{natbib}
\usepackage[utf8]{inputenc}	
\usepackage[T1]{fontenc}	
\usepackage{xcolor}		
\RequirePackage{doi}
\usepackage{hyperref}	
\usepackage{booktabs} 		
\usepackage{nicefrac}		
\usepackage{microtype}		
\usepackage{lineno}		
\usepackage{float}			
\usepackage{cleveref}

\newcommand{\isthesis}{0}

\usepackage{newfloat}
\DeclareFloatingEnvironment[name={Supplementary Figure}]{suppfigure}
\usepackage{sidecap}
\sidecaptionvpos{figure}{c}

\usepackage{titlesec}
\titlespacing\section{0pt}{12pt plus 3pt minus 3pt}{1pt plus 1pt minus 1pt}
\titlespacing\subsection{0pt}{10pt plus 3pt minus 3pt}{1pt plus 1pt minus 1pt}
\titlespacing\subsubsection{0pt}{8pt plus 3pt minus 3pt}{1pt plus 1pt minus 1pt}

\usepackage{tikz,xcolor,hyperref}

\definecolor{lime}{HTML}{A6CE39}
\DeclareRobustCommand{\orcidicon}{
	\begin{tikzpicture}
	\draw[lime, fill=lime] (0,0) 
	circle [radius=0.16] 
	node[white] {{\fontfamily{qag}\selectfont \tiny ID}};
	\draw[white, fill=white] (-0.0625,0.095) 
	circle [radius=0.007];
	\end{tikzpicture}
	\hspace{-2mm}
}
\foreach \x in {A, ..., Z}{\expandafter\xdef\csname orcid\x\endcsname{\noexpand\href{https://orcid.org/\csname orcidauthor\x\endcsname}
			{\noexpand\orcidicon}}
}

\newcommand{\sgn}{\operatorname{sgn}}
\newcommand{\atan}{\operatorname{atan}}

\title{Three-dimensional contrast transfer function in phase-contrast tomography}

\usepackage{xwatermark}
\newwatermark[firstpage,color=gray!60,angle=90,scale=0.32, xpos=-4.05in,ypos=0]{\href{https://doi.org/}{\color{gray}{Publication doi}}}
\newwatermark[firstpage,color=gray!60,angle=90,scale=0.32, xpos=3.9in,ypos=0]{\href{https://doi.org/10.48550/arXiv.2206.02688}{\color{gray}{Preprint doi}}}
\newwatermark[firstpage,color=gray!90,angle=0,scale=0.28, xpos=0in,ypos=-5in]{*correspondence: \texttt{darren.thompson@csiro.au}}

\usepackage{authblk}

\author[1,2,3\thanks{\tt{darren.thompson@csiro.au}}]{Darren A.~Thompson\orcidA{}}
\author[1]{Yakov I.~Nesterets\orcidB{}}
\author[4,5,3]{Konstantin M.~Pavlov\orcidC{}}
\author[6,5]{Timur E.~Gureyev\orcidD{}}

\affil[1]{Commonwealth Scientific and Industrial Research Organisation (CSIRO), Clayton, Victoria, Australia}
\affil[2]{Australian Synchrotron, Australian Nuclear Science and Technology Organisation (ANSTO), Clayton, Victoria, Australia}
\affil[3]{University of New England, Armidale, New South Wales, Australia}
\affil[4]{University of Canterbury, Christchurch, New Zealand}
\affil[5]{Monash University, Clayton, Victoria, Australia}
\affil[6]{The University of Melbourne, Parkville, Victoria, Australia}

\begin{document}

\twocolumn[ 
    \begin{@twocolumnfalse} 
  
\maketitle

\begin{abstract}
A new method is developed for three-dimensional (3D) reconstruction of multi-material objects using propagation-based X-ray phase-contrast tomography (PB-CT) with phase retrieval via the contrast transfer function (CTF) formalism. The approach differs from conventional PB-CT algorithms that apply phase retrieval on individual two-dimensional (2D) projections. Instead, this method involves performing phase retrieval to the CT-reconstructed volume in 3D. The CTF formalism is further extended to the cases of partially-coherent illumination and strongly absorbing samples. Simulated results demonstrate that the proposed post-reconstruction CTF method provides fast and stable phase retrieval, producing results equivalent to conventional pre-reconstruction 2D CTF phase retrieval. Moreover, it is shown that application can be highly localised to isolated objects of interest, without a significant loss of quality, thus leading to increased computational efficiency. Combined with the extended validity of the CTF to greater propagation distances, this method provides additional advantages over approaches based on the transport-of-intensity equation.
\end{abstract}

\vspace{0.35cm}

  \end{@twocolumnfalse} 
] 



\section{Introduction}

We recently proposed a method for three-dimensional (3D) reconstruction of the internal structure of objects in propagation-based X-ray phase-contrast tomography (PB-CT) \cite{Thompson2019FastTomography}. This method was based on the combination of conventional computed tomography (CT) with phase retrieval using the Transport of Intensity equation (TIE) \cite{Teague1983DeterministicSolution}. As such, the method was applicable to the so-called \textit{near-Fresnel} imaging conditions, i.e. the cases where the Fresnel number is much larger than unity: ${N_F} \equiv {a^2}/(\lambda R) >> 1$, where $a$ is the size of the imaged feature of interest, $\lambda$ is the radiation wavelength and $R$ is the effective free-space propagation distance between the imaged object and the detector \cite{Gureyev2004LinearRegion}. In the present paper, we extend this method to the whole of the Fresnel region, including the \textit{far-Fresnel} zone, where ${N_F} <  < 1$ . The latter imaging conditions can be encountered in practice, for example, in X-ray imaging \cite{Mayo2003X-rayMicrotomography} or electron microscopy \cite{Cowley1995DiffractionPhysics}. Mathematically, the principal difference between the method proposed in \cite{Thompson2019FastTomography} and the one developed in the present paper, can be explained in terms of the corresponding approaches to the linearization (with respect to the complex refractive index of the imaged object) of the general image intensity distribution expressed by the square modulus of the Fresnel diffraction integral. The method in \cite{Thompson2019FastTomography}, being based on the TIE, uses a linearization relying on the slow spatial variation of the refractive index. In contrast, the method developed in the present paper is based initially on the assumption of the weak scattering (first Born approximation), which assumes that the deviation of the refractive index from unity is small (see details in the next section). This approach is known as the contrast transfer function (CTF) or Fourier optics theory \cite{Cowley1995DiffractionPhysics, Pogany1997ContrastSource}. However, we subsequently show that, following the ideas described in \cite{Wu2003AImaging., Gureyev2004LinearRegion, Guigay2007MixedRegion., Nesterets2016PartiallyApproximation, Gureyev2017LinearConditions}, the two approaches can be merged, leading to a solution that is valid for refractive index distributions that can be represented as a sum of a slowly varying and a small component.

Considering another key aspect of the problem of the reconstruction of the 3D distribution of the complex refractive index inside an object from phase-contrast images collected at different incident illumination directions (or object orientations), we note that the conventional approach used for solution of this problem consists essentially of two stages. At the first stage, the collected 2D phase-contrast images are processed with the goal of recovering the complex amplitude in the object plane from the registered intensity distribution(s) \cite{Gureyev2004LinearRegion}, at each illumination direction. During the second stage, the distributions of the complex amplitude in the object planes, obtained at different illumination directions, are processed together to reconstruct the 3D distribution of the complex refractive index in the object by means of conventional CT techniques \cite{Natterer2001TheTomography}. Some related approaches exist \cite{Bronnikov1999ReconstructionTomography, Bronnikov2002TheoryTomography, Gureyev2006Phase-and-amplitudeTomography, Gureyev2022UnifiedTomography} that allow one to effectively merge these two stages into a single step, which may have advantages in terms of the computational efficiency and robustness.

In the two-stage PB-CT algorithms, typically, the phase retrieval is applied first, in 2D, at each illumination angle, followed by the 3D CT reconstruction. In contrast, in the 3D PB-CT methods described in \cite{Thompson2019FastTomography} and in the present paper, the 3D CT reconstruction is effectively applied to the raw phase-contrast images first and the phase retrieval is applied in 3D after that, even though the two operations appear as parts of a single analytical expression. As explained in \cite{Thompson2019FastTomography}, the latter methodology can be substantially advantageous e.g. in the case of objects containing several distinct components, each spatially localized to a 3D area ${\Omega _m}$, with different locally-constant ratios ${\gamma _m} \equiv \Delta ({\bf{r}})/\beta ({\bf{r}}),\,\,\,{\bf{r}} \in {\Omega _m}$, of the real decrement and the imaginary part of the complex refractive index $n({\bf{r}}) = 1 - \Delta ({\bf{r}}) + i\beta ({\bf{r}})$ \cite{Beltran20102DDistance}. In this case, the computationally expensive 3D CT reconstruction step can be performed only once for the whole object volume, followed by repeated phase retrieval operations localized to different (smaller) 3D areas ${\Omega _m}$. A highly efficient and stable 3D phase retrieval method based on the monomorphous (homogeneous) TIE (TIE-Hom) \cite{Paganin2002SimultaneousObject} can be applied here locally in ${\Omega _m}$ with the constant value ${\gamma _m}$. The ability to apply phase retrieval to the CT-reconstructed volume or to localized sub-volumes corresponding to different objects of interest, provides useful flexibility to PB-CT processing workflows as demonstrated below \cite{Donato2022OptimizationTomography, Pollock2023RobustObjectsArxiv}.

As will be described in this paper, a very similar methodology to the one proposed in \cite{Thompson2019FastTomography} can be implemented, replacing the use of the transport-of-intensity equation with a CTF equivalent. In \cref{s3:derivation_1} the 3D CTF formalism is derived for weakly scattering objects. Using this formalism, \cref{s3:derivation_2}  presents reconstruction formulae for the complex refractive index, for both general and monomorphous objects. In \cref{s3:derivation_3}, we further generalise these results to the cases of strongly absorbing objects and partially-coherent illumination. Numerically simulated examples of application of the proposed CT reconstruction techniques can be found in \cref{s3:numerical_simulation_framework}. Finally, we summarize the main results of the paper, together with the relevant validity conditions, in \cref{s3:conclusion}

\section{3D CTF for weakly scattering objects} \label{s3:derivation_1}

\begin{figure}[!htbp]
    \centering
        \includegraphics[width=\linewidth]{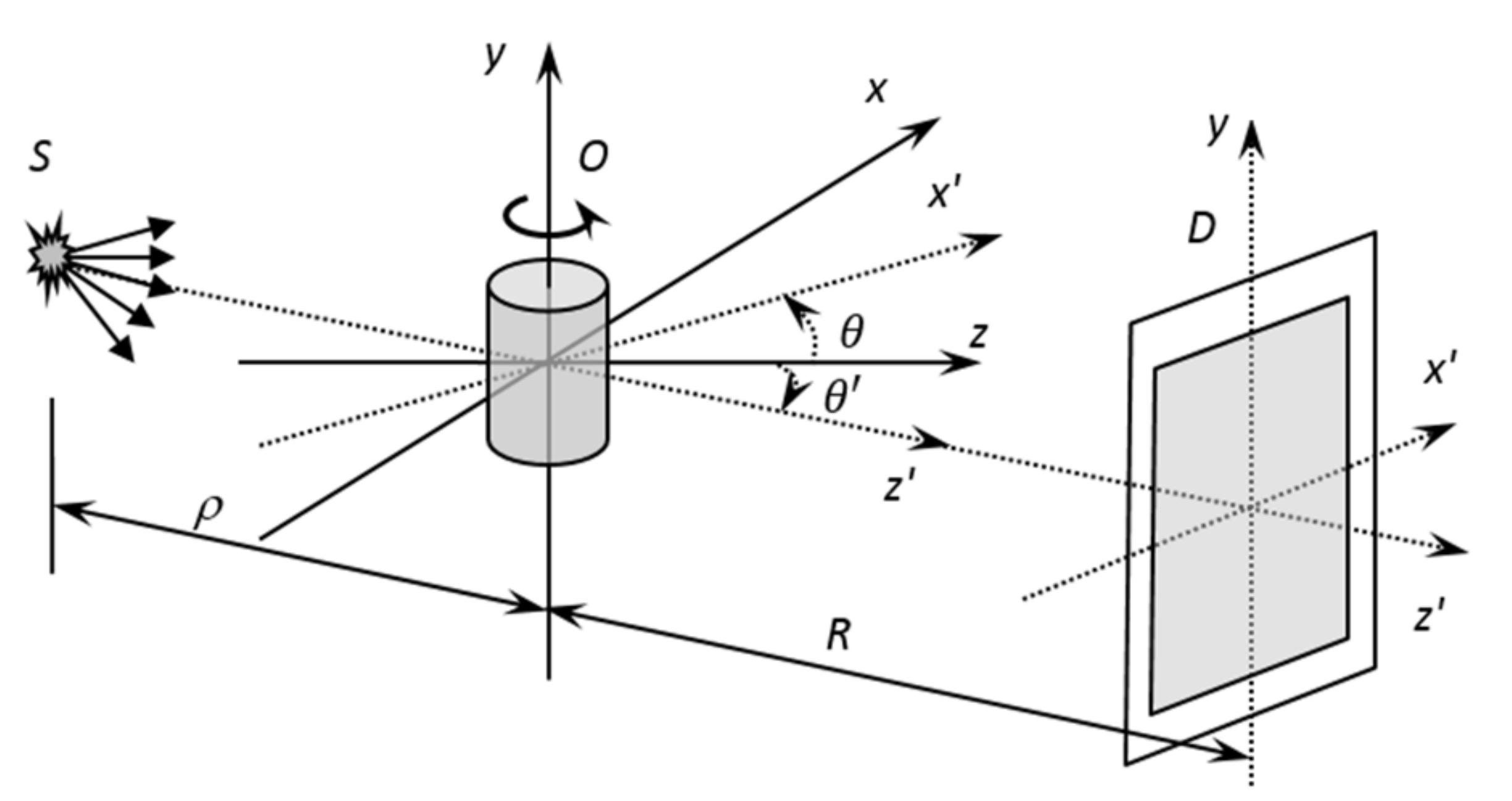}%
        \caption{PB-CT experimental setup.}
        \label{fig3:pbct_experimental_setup_ctf}
\end{figure}

Consider the PB-CT imaging system schematically shown in \cref{fig3:pbct_experimental_setup_ctf}. In the following, we assume that the dimensions of the object are small compared to the source-to-object distance $\rho$  and $\rho  \gg R$, the propagation distance, i.e. that the incident wave is planar.  Let an object be illuminated by a monochromatic plane X-ray wave with wavelength $\lambda$ and intensity ${I_{in}}$, $I_{in}^{1/2}\exp (ikz')$ with $k = 2\pi /\lambda $.  The direction $z'$ of the incident X-ray wave forms an angle $\theta '$ with the z axis of the objects coordinate system $(x,y,z)$, $ - \pi /2 \le \theta ' < \pi /2$ , and $\theta  = \theta ' + \pi /2$.  The phase-contrast image of the object is recorded on a position-sensitive detector located at a distance $R$ downstream from the object. Interactions of the X-rays and object matter are described via the spatial distribution of the complex refractive index.

Also assumed is the interaction of the incident X-rays and the object are accurately described by the complex scalar transmission function, 
\begin{equation}
    \label{eq3:complex_scalar_transmission}
    {q_\theta }\left( {x',y} \right) = \exp\left[ { - {B_\theta }(x',y) + i{\varphi _\theta }(x',y)} \right],
\end{equation}
consisting of the amplitude attenuation ${B_\theta }(x',y) = k{P_\theta }\beta \left( {x',y} \right)$ and phase ${\varphi _\theta }(x',y) =  - k{P_\theta }\Delta \left( {x',y} \right)$ functions, both of which are defined in terms of the projection operator, 
\begin{multline}
    \label{eq3:projection_operator}
    {P_\theta }f\left( {x',y} \right) = \\
    \mathop \smallint \limits_{ - \infty }^\infty  \mathop \smallint \limits_{ - \infty }^\infty  f\left( {x,y,z} \right)\delta \left( {x' - x\;\sin \theta  - z\;\cos \theta } \right)dxdz,
\end{multline}
where $\delta$ represents the Dirac delta function. 
The evolution of a paraxial transmitted wave in the free half-space $z' > 0$ can be described by the 2D Fresnel diffraction integral \cite{Goodman1996IntroductionOptics},
\begin{multline}
    \label{eq3:paraxial_wave}
    \psi _\theta ^R(x',y) = I_{in}^{1/2}\frac{{\exp \left( {ikR} \right)}}{{i\lambda R}}\\
    \times \int\limits_{ - \infty }^\infty  {\int\limits_{ - \infty }^\infty  {\exp \left\{ {\frac{{i\pi }}{{\lambda R}}\left[ {{{\left( {x' - x''} \right)}^2} + {{(y - y'')}^2}} \right]} \right\}}}\\
    \times {q_\theta }\left( {x'',y''} \right)dx''dy''.
\end{multline}
Utilising the 2D Fresnel free-space propagator,
\begin{equation}
    \label{eq3:fresnel_propagator_2d}
     P_2^R\left( {x',y} \right) = {\left( {i\lambda R} \right)^{ - 1}}\exp \left[ {i\pi \left( {{{x'}^2} + {y^2}} \right)/(\lambda R)} \right].
\end{equation}
\Cref{eq3:paraxial_wave} can be conveniently presented as a 2D convolution, 
\begin{equation}
    \label{eq3:paraxial_wave_conv}
    \psi _\theta ^R(x',y) = I_{in}^{1/2} \exp (ikR)\left({P_2^R * {q_\theta }}\right)(x',y).       
\end{equation}
The 2D convolution of two functions is defined as:
\begin{equation}
    \label{eq3:convolution_2d}
    (f * g)\left( {x,y} \right) = \int\limits_{ - \infty }^\infty  {\int\limits_{ - \infty }^\infty  {f(x',y')\,g(x - x',y - y')\,dx'dy'} } .
\end{equation}
The spatial distribution of the propagated intensity in the detector plane is thus,
\begin{equation}
    \label{eq3:intensity_r}
    I_\theta ^R\left( {x',y} \right) = {\left| {\psi _\theta ^R(x',y)} \right|^2}.
\end{equation}
The Fourier transform of the image intensity is given by \cite{Guigay1977FourierHolograms},
\begin{multline}     
\label{eq3:intensity_r_ctf}
{F_2}I_\theta ^R\left( {\xi ',\eta } \right) = I_{in} \mathop \smallint \limits_{ - \infty }^\infty  \mathop \smallint \limits_{ - \infty }^\infty  \exp \left[ { - i2\pi \left( {x'\xi ' + y\eta } \right)} \right]\\
 \times {q_\theta }\left( {x' - \frac{1}{2} \lambda R\xi ',y -\frac{1}{2} \lambda R\eta } \right)\\
 \times q_\theta ^ * \left({x' + \frac{1}{2} \lambda R\xi ',y + \frac{1}{2} \lambda R\eta }\right)dx'dy,
\end{multline}
where the superscript asterisk denotes the complex conjugate and the 2D Fourier transform is defined as follows\ifthenelse{\equal{\isthesis}{1}}{ (note that the definition of the Fourier transform differs to that used in \cref{chapter:tie_hom_3d})},
\begin{equation}
    \label{eq3:fourier_2d}
    {F_2}g\left( {\xi ',\eta } \right) = \mathop \smallint \limits_{ - \infty }^\infty  \mathop \smallint \limits_{ - \infty }^\infty  \exp \left[ { - i2\pi \left( {x'\xi ' + y\eta } \right)} \right]g\left( {x',y} \right)dx'dy.
\end{equation}
The phase retrieval problem is that of finding a solution to the non-linear expressions of \cref{eq3:intensity_r,eq3:intensity_r_ctf} with respect to ${q_\theta }$. To simplify this task, linearized approximations are generally used, for which many methods have been derived \cite{Cowley1995DiffractionPhysics, Guigay1977FourierHolograms, Teague1983DeterministicSolution, Pogany1997ContrastSource, Paganin2002SimultaneousObject, Gureyev2003CompositeRegion, Wu2003AImaging., Gureyev2004LinearRegion, Gureyev2006LinearIllumination}.

One linearization approach, originally developed by Guigay \cite{Guigay1977FourierHolograms}, can be derived under the assumptions of weak object attenuation and weak or slowly-varying phase:
\begin{equation}
    \label{eq3:weak_attenuation_phase}
    \begin{array}{l}
        {B_\theta }(x',y) \ll 1,\\
        \left| {{\varphi _\theta }(x' - \lambda R\xi ',y - \lambda R\eta ) - {\varphi _\theta }\left( {x',y} \right)} \right| \ll 1.
    \end{array}
\end{equation}
Applying these assumptions in addition to a uniform incident intensity distribution ${I_{in}}$ in the object plane, Guigay obtained the linearized expression for the propagated intensity distribution (extended to the 2D case), \textit{cf}. \cite{Guigay1977FourierHolograms}
\begin{multline}
    \label{eq3:intensity_r_ctf_weak}
    {F_2}I_\theta ^R \left( {\xi ',\eta } \right) \cong {I_{in}}\\
    \times \left\{ \delta \left( {\xi ',\eta } \right) - 2{F_2}{B_\theta }\left( {\xi ',\eta } \right)\cos \left[ {\pi \lambda R\left( {{{\xi '}^2} + {\eta ^2}} \right)} \right]\right.\\
    \left. + 2{F_2}{\varphi _\theta }\left( {\xi ',\eta } \right)\sin \left[ {\pi \lambda R\left( {{{\xi '}^2} + {\eta ^2}} \right)} \right]\right\}.
\end{multline}
Recall that the mathematical basis for CT can be described by the inversion equation \cite{Natterer2001TheTomography},
\begin{equation}
    \label{eq3:ct_operator}
    \Re {F_2}{P_\theta }f(x,y,z) = f(x,y,z),
\end{equation}
where $\Re $ is the filtered back-projection (FBP) operator,
\begin{multline}
    \label{eq3:fbp_operator}
    \Re h\left( {x,y,z} \right) = \\{\rm{\;}}\mathop \smallint \limits_0^\pi  \mathop \smallint \limits_{ - \infty }^\infty  \mathop \smallint \limits_{ - \infty }^\infty  \exp \left\{ {i2\pi \left[ {\xi '\left( {x{\rm{\;}}\sin \theta  + z{\rm{\;}}\cos \theta } \right) + \eta y} \right]} \right\} \\
    \times h\left( {\xi ',\eta ,\theta } \right)\left| {\xi '} \right|d\xi 'd\eta d\theta .
\end{multline}
A generalised form of the 2D Fourier derivative theorem \cite{Paganin2006CoherentOptics} is given by,
\begin{equation}
    \label{eq3:fourier_derivative_theorem}
    {F_2}\left[ {\frac{{{\partial ^m}}}{{\partial {{x'}^m}}}\frac{{{\partial ^n}}}{{\partial {y^n}}}g(x',y)} \right](\xi ',\eta ) = {\left( {2\pi i\xi '} \right)^m}{\left( {2\pi i\eta } \right)^n}{F_2}g(\xi ',\eta ).
\end{equation}
Using \cref{eq3:fourier_derivative_theorem}, an explicit form of the Fourier transform of the 2D Laplacian of a function in the detector plane can thus be expressed as the identity, 
\begin{equation}
    \label{eq3:fourier_laplacian_identity}
    {F_2}\nabla _ \bot ^2g(\xi ',\eta ) =  - 4{\pi ^2}\left( {{{\xi '}^2} + {\eta ^2}} \right){F_2}g(\xi ',\eta ),
\end{equation}
where ${\nabla _ \bot } = \left( {\frac{\partial }{{\partial x'}},\frac{\partial }{{\partial y}}} \right)$.
Moreover, when the 3D Laplacian is applied to both sides of \cref{eq3:ct_operator}, the following equality can be established,
\begin{equation}
    \label{eq3:fourier_ct_laplacian}
    \Re {F_2}\nabla _ \bot ^2{P_\theta }f(x,y,z) = {\nabla ^2}f(x,y,z).
\end{equation}
Here, $\nabla  = \left( {\frac{\partial }{{\partial x}},\frac{\partial }{{\partial y}},\frac{\partial }{{\partial z}}} \right)$.

Additionally, \cref{eq3:fourier_laplacian_identity,eq3:fourier_ct_laplacian} can be generalised for any positive integer power $n$ of the Laplacian,
\begin{equation}
    \label{eq3:fourier_laplacian_identity_n}
    {F_2}\nabla _ \bot ^{2n}g(\xi ',\eta ) = {\left[ { - 4{\pi ^2}\left( {{{\xi '}^2} + {\eta ^2}} \right)} \right]^n}{F_2}g(\xi ',\eta ),
\end{equation}
\begin{equation}
    \label{eq3:fourier_ct_laplacian_n}
    \Re {F_2}\nabla _ \bot ^{2n}{P_\theta }f(x,y,z) = \nabla^{2n}f(x,y,z).
\end{equation}
Importantly, this last property, \cref{eq3:fourier_ct_laplacian,eq3:fourier_ct_laplacian_n}, demonstrates the ability to switch the order of differentiation and filtered back-projection, with the corresponding change of the dimensionality of the Laplacian operator between 2D projections and the reconstructed volume in 3D real space. A post-CT reconstruction method for phase retrieval of monomorphous objects in the TIE regime has been derived \cite{Thompson2019FastTomography} exploiting this property.

Below, we adapt this approach for the CTF approximation by expressing the Fresnel propagator components in terms of a Laplacian to utilise the relationship in \cref{eq3:fourier_ct_laplacian_n}.
Let ${K_{R,\theta }}\left( {x',y} \right) \buildrel \Delta \over = 1 - I_\theta ^R\left( {x',y} \right)/{I_{in}}$ designate the in-line contrast function, with ${F_2}P_2^a\left( {\xi ',\eta } \right) = \cos \left[ {\pi \lambda R\left( {{{\xi '}^2} + {\eta ^2}} \right)} \right]$ and ${F_2}P_2^p\left( {\xi ',\eta } \right) =  - \sin \left[ {\pi \lambda R\left( {{{\xi '}^2} + {\eta ^2}} \right)} \right]$  being the 2D Fourier transforms of the amplitude and phase Fresnel propagators respectively \cite{Nesterets2016PartiallyApproximation}. Rearranging \cref{eq3:intensity_r_ctf_weak} produces the expression,
\begin{multline}
    \label{eq3:ctf_contrast}
    {F_2}{K_{R,\theta }}\left( {\xi ',\eta } \right) =\\
    2{F_2}{B_\theta }\left( {\xi ',\eta } \right){F_2}P_2^a\left( {\xi ',\eta } \right) + 
    2{F_2}{\varphi _\theta }\left( {\xi ',\eta } \right){F_2}P_2^p\left( {\xi ',\eta } \right).
\end{multline}
The 2D Fourier transform of the amplitude Fresnel propagator, ${F_2}P_2^a\left( {\xi ',\eta } \right)$, when expressed as a Taylor series has the form,
\begin{equation}
    \label{eq3:amplitude_propagator_taylor}
    {F_2}P_2^a\left( {\xi ',\eta } \right) = \sum\limits_{n = 0}^\infty  {\frac{{{{\left( { - 1} \right)}^n}}}{{\left( {2n} \right)!}}} {\left[ {\pi \lambda R\left( {{{\xi '}^2} + {\eta ^2}} \right)} \right]^{2n}}.
\end{equation}
Multiplying the expansion by the Fourier transform of the 2D projection ${F_2}{P_\theta }f\left( {\xi ',\eta } \right)$ leads to, 
\begin{multline}
    \label{eq3:amplitude_projectionr_taylor}
    {F_2}P_2^a\left( {\xi ',\eta } \right){F_2}{P_\theta }f\left( {\xi ',\eta } \right) = \\
    \sum\limits_{n = 0}^\infty  {\frac{{{{\left( { - 1} \right)}^n}}}{{\left( {2n} \right)!}}} {\left[ {\pi \lambda R\left( {{{\xi '}^2} + {\eta ^2}} \right)} \right]^{2n}}{F_2}{P_\theta }f\left( {\xi ',\eta } \right).
\end{multline}
Using \cref{eq3:fourier_laplacian_identity_n}, the right-hand side of \cref{eq3:amplitude_projectionr_taylor} can be re-written in terms of the power series of the Laplacian of projections, 
\begin{multline}
    \label{eq3:amplitude_propagator_laplacian_n}
    {F_2}P_2^a\left( {\xi ',\eta } \right){F_2}{P_\theta }f\left( {\xi ',\eta } \right) = \\\sum\limits_{n = 0}^\infty  {C_n^a{F_2}\left\{ {{{\left[ {\nabla _ \bot ^2} \right]}^{2n}}{P_\theta }f} \right\}\left( {\xi ',\eta } \right)} ,
\end{multline}
where $C_n^a = \frac{{{{\left( { - 1} \right)}^n}}}{{\left( {2n} \right)!}}{\left[ {\frac{{\lambda R}}{{4\pi }}} \right]^{2n}}$.

It is worth noting that according to \cref{eq3:amplitude_propagator_laplacian_n} the amplitude contrast, corresponding to the first term in the right-hand side of \cref{eq3:ctf_contrast}, can be expressed as a power series of the Laplacian of the attenuation function,
\[(P_2^a * {B_\theta })\left( {x',y} \right) = \sum\limits_{n = 0}^\infty  {C_n^a{{\left[ {\nabla _ \bot ^2} \right]}^{2n}}{B_\theta }(x',y)} .\]
Similarly, the phase contrast, corresponding to the second term in the right-hand side of \cref{eq3:ctf_contrast}, can be expressed as a power series of the Laplacian of the phase function,
\[(P_2^p * {\varphi _\theta })\left( {x',y} \right) = \sum\limits_{n = 0}^\infty  {C_n^p{{\left[ {\nabla _ \bot ^2} \right]}^{2n + 1}}{\varphi _\theta }(x',y)} ,\]
where $C_n^p = \frac{{{{\left( { - 1} \right)}^n}}}{{\left( {2n + 1} \right)!}}{\left[ {\frac{{\lambda R}}{{4\pi }}} \right]^{2n + 1}}$.

Using \cref{eq3:fourier_ct_laplacian_n}, the filtered back-projection operator applied to the right-hand side of \cref{eq3:amplitude_propagator_laplacian_n} is expressed as follows,
\begin{multline}
    \Re \left( {\sum\limits_{n = 0}^\infty  {C_n^a{F_2}\left\{ {{{\left[ {\nabla _ \bot ^2} \right]}^{2n}}{P_\theta }f} \right\}} } \right)\left( {x,y,z} \right) \\
    =\sum\limits_{n = 0}^\infty  {C_n^a\Re \left( {{F_2}{{\left[ {\nabla _ \bot ^2} \right]}^{2n}}{P_\theta }f} \right)(x,y,z)} \\
    = \sum\limits_{n = 0}^\infty  {C_n^a{{\left[ {{\nabla ^2}} \right]}^{2n}}} f(x,y,z).
\end{multline}

Hence,
\begin{equation}
    \label{eq3:fbp_amplitude}
    \begin{split}       
    \Re \left( {{F_2}P_2^a{F_2}{P_\theta }f} \right)\left( {x,y,z} \right) & = \sum\limits_{n = 0}^\infty  {C_n^a{{\left[ {{\nabla ^2}} \right]}^{2n}}} f(x,y,z) \\ 
    & = (P_3^{a,R} * f)\left( {x,y,z} \right),
    \end{split}    
\end{equation}
where we introduced a 3D amplitude propagator, $P_3^{a,R}(x,y,z)$.

An equivalent treatment can be similarly applied to the phase component of the Fresnel propagator,  ${F_2}P_2^p\left( {\xi ',\eta } \right) =  - \sin \left[ {\pi \lambda R\left( {{{\xi '}^2} + {\eta ^2}} \right)} \right]$, resulting in the equation,
\begin{multline}
    \label{eq3:phase_propagator_laplacian_n}
    {F_2}P_2^p\left( {\xi ',\eta } \right){F_2}{P_\theta }f\left( {\xi ',\eta } \right) = \\\sum\limits_{n = 0}^\infty  {C_n^p{F_2}\left\{ {{{\left[ {\nabla _ \bot ^2} \right]}^{2n + 1}}{P_\theta }f} \right\}\left( {\xi ',\eta } \right)}.
\end{multline}

Hence,
\begin{multline}
    \label{eq3:fbp_phase}
    \Re \left( {{F_2}P_2^p{F_2}{P_\theta }f} \right)\left( {x,y,z} \right)
    = \sum\limits_{n = 0}^\infty  {C_n^p{{\left[ {{\nabla ^2}} \right]}^{2n + 1}}} f(x,y,z) \\
    = (P_3^{p,R} * f)\left( {x,y,z} \right),
\end{multline}
where we introduced a 3D amplitude propagator, $P_3^{p,R}(x,y,z)$.

Applying the amplitude and phase reconstructions given by \cref{eq3:fbp_amplitude,eq3:fbp_phase} to \cref{eq3:ctf_contrast}, with proper substitutions for $\beta$ and $\Delta$, and expressing the result in 3D Fourier space results in,
\begin{multline}
    \label{eq3:fbp_ctf_3d}
    {F_3}\Re {F_2}{K_{R,\theta }}\left( {\xi ,\eta ,\zeta } \right) = 2k{F_3}\beta \left( {\xi ,\eta ,\zeta } \right){F_3}P_3^{a,R}\left( {\xi ,\eta ,\zeta } \right) \\
    - 2k{F_3}\Delta \left( {\xi ,\eta ,\zeta } \right){F_3}P_3^{p,R}\left( {\xi ,\eta ,\zeta } \right).
\end{multline}
Here, 
\begin{equation}
    \label{eq3:amplitude_propagator}
    {F_3}P_3^{a,R}\left( {\xi ,\eta ,\zeta } \right) = \cos \left[ {\pi \lambda R\left( {{\xi ^2} + {\eta ^2} + {\zeta ^2}} \right)} \right],    
\end{equation}
\begin{equation}
    \label{eq3:phase_propagator}
    {F_3}P_3^{p,R}\left( {\xi ,\eta ,\zeta } \right) =  - \sin \left[ {\pi \lambda R\left( {{\xi ^2} + {\eta ^2} + {\zeta ^2}} \right)} \right],
\end{equation}
are the 3D Fourier transforms of the 3D amplitude and phase Fresnel propagators, respectively, and ${F_3}$ represents the 3D Fourier transform operator,
\begin{multline}
    \label{eq3:fourier_operator_3d}
    {F_3}g\left( {\xi ,\eta ,\zeta } \right) = \\\mathop \smallint \limits_{ - \infty }^\infty  \mathop \smallint \limits_{ - \infty }^\infty  \mathop \smallint \limits_{ - \infty }^\infty  \exp \left[ { - i2\pi \left( {x\xi  + y\eta  + z\zeta } \right)} \right]g\left( {x,y,z} \right)dxdydz.
\end{multline}

\section{Application of 3D CTFs for PB-CT reconstruction} \label{s3:derivation_2}
Examination of the similarity between \cref{eq3:ctf_contrast,eq3:fbp_ctf_3d} establishes an important link between 2D and 3D expressions for propagation-based phase contrast in the CTF regime. \cref{eq3:fbp_ctf_3d} represents a generalized form of 3D CTF phase contrast that provides a basis for phase retrieval on CT reconstructed images.

Moreover, with some algebraic manipulation, general solutions for the Fourier transforms of both $\beta$ and $\Delta$ can be obtained for two propagation distances ${R_1}$ and ${R_2}$ (the argument list $(\xi ,\eta ,\varsigma )$, has been omitted for brevity):
\begin{equation}
    \label{eq3:beta_fourier_3d}
    {F_3}\beta  = \frac{{{\Im _{{R_2}}}\;{F_3}P_3^{p,{R_1}} - {\Im _{{R_1}}}\;{F_3}P_3^{p,{R_2}}}}{{2k{F_3}P_3^{p,{R_1} - {R_2}}}},
\end{equation}
\begin{equation}
    \label{eq3:delta_fourier_3d}
    {F_3}\Delta  = \frac{{ - {\Im _{{R_1}}}\,{F_3}P_3^{a,{R_2}} + {\Im _{{R_2}}}\,{F_3}P_3^{a,{R_1}}}}{{2k{F_3}P_3^{p,{R_1} - {R_2}}}}.
\end{equation}
Here, ${\Im _R}\left( {\xi ,\eta ,\zeta } \right) = {F_3}\Re {F_2}{K_{R,\theta }}\left( {\xi ,\eta ,\zeta } \right)$ represents the 3D Fourier transform of the reconstructed in-line contrast function at the propagation distance $R$, see \cref{eq3:fbp_ctf_3d}. Additionally, ${F_3}P_3^{p,{R_1} - {R_2}} = {F_3}P_3^{p,{R_1}}{F_3}P_3^{a,{R_2}} - {F_3}P_3^{p,{R_2}}{F_3}P_3^{a,{R_1}}$.

The structure of \cref{eq3:beta_fourier_3d,eq3:delta_fourier_3d} illustrates that both the real and imaginary parts of the refractive index decrement in the object plane can be retrieved by subtracting weighted 3D Fourier-filtered FBP reconstructions of the propagated intensity at two distances. Importantly, as the ${\Im _R}\left( {\xi ,\eta ,\zeta } \right)$ terms are decoupled from the associated propagator terms, absorption/phase retrieval can be performed after conventional FBP-CT reconstructions. Furthermore, this decoupling also implies that retrieval can be applied to a localized CT-reconstructed sub-volume as the applied 3D propagator functions are not bound or constrained by ${\Im _R}\left( {\xi ,\eta ,\zeta }\right)$.

Assume that the object is monomorphous, such that a spatially independent (but energy-dependent) proportionality constant, $\gamma  = \Delta /\beta $, holds for the complex refractive index \cite{Paganin2002SimultaneousObject, Mayo2003X-rayMicrotomography}. This assumption is valid, for example, for objects consisting of a single material and objects composed of light elements (with atomic numbers $Z < 10$) when irradiated with high-energy X-rays (60-500 keV)\cite{Wu2003AImaging.}.

Applying this property to the generalised CTF expression in \cref{eq3:fbp_ctf_3d} provides a monomorphous form of the CTF, denoted CTF-Hom, allowing for the reconstruction of the object’s linear attenuation coefficient, $\mu(x,y,z)=2k \beta(x,y,z) $, from a single set of intensity measurements collected at a single propagation distance $R$,

\begin{equation}
\begin{aligned}
    \label{eq3:ctf_hom_mu_3d}
    {F_3}\mu \left( {\xi ,\eta ,\zeta } \right) &= {\frac{{\Im _R}\left( {\xi ,\eta ,\zeta } \right)}{{F_3}P_3^{a,R}\left( {\xi ,\eta ,\zeta } \right) - \gamma {F_3}P_3^{p,R}\left( {\xi ,\eta ,\zeta } \right)}}\\
    &= {\frac{\sgn \left(\gamma\right) {\Im _R}\left( {\xi ,\eta ,\zeta } \right)}{\sqrt{\gamma^2 + 1}\sin{\left[\pi\lambda R u^2 + \atan{\left(\gamma^{-1}\right)}\right]}}},
\end{aligned}
\end{equation}
where $u^2 = \xi^2 + \eta^2 + \zeta^2$. If required, $\beta(x,y,z)$ and $\Delta(x,y,z)$ can be easily recovered from $\mu(x,y,z)$.

Moreover, it can be shown that the CTF-Hom reconstruction formula, \cref{eq3:ctf_hom_mu_3d}, reduces to the corresponding TIE form under the assumption that the complex transmission function, ${q_\theta }\left( {x',y} \right)$ is slowly varying on the length scale $\sqrt {\lambda R}$ at all $\theta $ whereby the in-line contrast function ${K_{R,\theta }}\left( {x',y} \right)$ is band-limited to the spectral region ${\xi '^2} + {\eta ^2} \ll {(\lambda R)^{ - 1}}$.  Applying these constraints to \cref{eq3:ctf_hom_mu_3d} results in the cosine function in \cref{eq3:amplitude_propagator} reducing to 1 and the sine function in \cref{eq3:phase_propagator} being replaced by its argument, giving the expression equivalent to that derived in \cite{Thompson2019FastTomography},

\begin{equation}
    \label{eq3:tie_hom_mu_3d}
    {F_3}\mu \left( {\xi ,\eta ,\zeta } \right) = \frac{{\Im _R}{\left( {\xi ,\eta ,\zeta } \right)}}{{1 + \pi \gamma \lambda R\left( {{\xi ^2} + {\eta ^2} + {\zeta ^2}} \right)}}.
\end{equation}

In its original form, the CTF approximation as derived above does not impose constraints upon propagation distance as TIE approximations do. However, it is quite restrictive due to the assumption of weak absorption, limiting its usability in real-life imaging applications.
Subsequent works have extended the validity of the CTF by way of a slowly-varying object approximation, allowing it to be used for absorbing objects \cite{Wu2003AImaging., Gureyev2004LinearRegion, Guigay2007MixedRegion., Nesterets2016PartiallyApproximation, Gureyev2017LinearConditions}. 

\section{Extension to partially-coherent illumination and strongly absorbing objects} \label{s3:derivation_3}

Nesterets and Gureyev \cite{Nesterets2016PartiallyApproximation} developed a CTF approach for partially coherent illumination, with validity extended to strongly absorbing objects. An equivalent approach can be applied to extend the validity of 3D CTF.

Let the attenuation function be the sum of a small (\textit{sm}), and a slowly varying (\textit{sl}) components:
\begin{equation}
    \label{eq3:attenuation_sm_sl}
    {B_\theta }(x',y) = {B_{\theta ,sm}}(x',y) + {B_{\theta ,sl}}(x',y).
\end{equation}
It is assumed that the small component satisfies $\left| {{B_{\theta ,sm}}(x',y)} \right| \ll 1$ and   ${B_{\theta ,sl}}(x',y)$ varies slowly on the length scale relative to the width of the partially-coherent free-space propagator \cite{Nesterets2016PartiallyApproximation}. Introducing the complex function,
\begin{equation}
    \label{eq3:complex_refraction_sm}
    \Phi (x',y) =  - {B_{\theta ,sm}}(x',y) + i{\varphi _\theta }(x',y),
\end{equation}
and the contact intensity of the slowly varying attenuation component,
\begin{equation}
    \label{eq3:intensity_sl}
    {I_{\theta ,sl}}\left( {x',y} \right) = \exp \left[ { - 2{B_{\theta ,sl}}(x',y)} \right],
\end{equation}
the combined propagated intensity can be approximated as in \cite{Nesterets2016PartiallyApproximation}:
\begin{equation}
    \label{eq3:intensity_sl_sm}
    I_\theta ^R\left( {x',y} \right) \cong {I_{in}}{I_{\theta ,sl}}\left( {x',y} \right)\left[ {1 + 2{\mathop{\rm Re}\nolimits} \left( {\Phi  * \tilde P_2^R} \right)\left( {x',y} \right)} \right].
\end{equation}
$\tilde P_2^R\left( {x',y} \right) = \left( {P_2^R * {P_{sys}}} \right)\left( {x',y} \right)$ is the 2D partially-coherent Fresnel free-space propagator obtained by convolving the fully-coherent Fresnel free-space propagator with the point-spread function of the imaging system, ${P_{sys}}\left( {x',y} \right)$. The latter takes into account partial coherence of the incident illumination, as well as finite resolution of the detector \cite{Nesterets2016PartiallyApproximation}.

Rearranging \cref{eq3:intensity_sl_sm} by bringing the incident intensity to the left-hand side and applying the negative logarithm to both sides results in an expression representing the propagated contrast function: 
\begin{multline}
    \label{eq3:contrast_sl_sm}
    {\tilde K_{R,\theta }}\left( {x',y} \right) \cong \\2{B_{\theta ,sl}}(x',y) - \log \left[ {1 + 2{\mathop{\rm Re}\nolimits} \left( {\Phi  * \tilde P_2^R} \right)\left( {x',y} \right)} \right],
\end{multline}
with ${\tilde K_{R,\theta }}\left( {x',y} \right) =  - \log \left( {\frac{{I_\theta ^R\left( {x',y} \right)}}{{{I_{in}}}}} \right).$
The logarithm on the right-hand side of \cref{eq3:contrast_sl_sm} can be linearized using the approximation $\log (1 + x) \cong x$ as the contrast due to weak absorption and phase contrast is assumed to be small relative to unity,
\begin{multline}
    \label{eq3:contrast_ctf_partial_coherent}
    {\tilde K_{R,\theta }}\left( {x',y} \right) \cong \\2{B_{\theta ,sl}}(x',y) + 2\left( {{B_{\theta ,sm}} * \tilde P_2^{a,R}} \right)\left( {x',y} \right) \\+ 2\left( {{\varphi _\theta } * \tilde P_2^{p,R}} \right)\left( {x',y} \right),
\end{multline}
where $\tilde P_2^{a,R}\left( {x',y} \right) = \left( {P_2^{a,R} * {P_{sys}}} \right)\left( {x',y} \right)$ and   $\tilde P_2^{p,R}\left( {x',y} \right) = \left( {P_2^{p,R} * {P_{sys}}} \right)\left( {x',y} \right)$ are the 2D partially-coherent amplitude and phase Fresnel propagators, respectively.

\Cref{eq3:contrast_ctf_partial_coherent} represents a CTF approximation that is applicable and valid for strongly absorbing (but slowly varying) objects containing weakly absorbing features. Moreover, due to the above assumption that ${B_{\theta ,sl}}$ varies slowly on the length scale relative to the width of the partially-coherent free-space propagator, the first two terms in the right-hand side of \cref{eq3:contrast_ctf_partial_coherent} may be combined, such that ${B_{\theta ,sl}}(x',y) + \left( {{B_{\theta ,sm}} * \tilde P_2^{a,R}} \right)\left( {x',y} \right) \cong \left( {{B_\theta } * \tilde P_2^{a,R}} \right)\left( {x',y} \right)$. 

Applying this simplification into \cref{eq3:contrast_ctf_partial_coherent} results in,
\begin{equation}
    \label{eq3:contrast_ctf_approx}
    {\tilde K_{R,\theta }}\left( {x',y} \right) \cong 2\left( {{B_\theta } * \tilde P_2^{a,R}} \right)\left( {x',y} \right) + 2\left( {{\varphi _\theta } * \tilde P_2^{p,R}} \right)\left( {x',y} \right),
\end{equation}
which is equivalent to the CTF approximation given by \cref{eq3:ctf_contrast}.
Applying the same 3D treatment to \cref{eq3:contrast_ctf_approx} as was done with \cref{eq3:ctf_contrast} to derive \cref{eq3:fbp_ctf_3d} provides an expression for the FBP reconstructed 3D contrast function, which is similar to \cref{eq3:fbp_ctf_3d},
\begin{multline}   
    \label{eq3:fbp_ctf_3d_partial_coherent}
    {F_3}\Re {F_2}{\tilde K_{R,\theta }}\left( {\xi ,\eta ,\zeta } \right) \cong 2k{F_3}\beta \left( {\xi ,\eta ,\zeta } \right){F_3}\tilde P_3^{a,R}\left( {\xi ,\eta ,\zeta } \right) \\
    - 2k{F_3}\Delta \left( {\xi ,\eta ,\zeta } \right){F_3}\tilde P_3^{p,R}\left( {\xi ,\eta ,\zeta } \right).
\end{multline}
Here, 
\begin{align*}
\tilde P_3^{a,R}\left( {x,y,z} \right) & = \left( {P_3^{a,R} * {P_{3,sys}}} \right)\left( {x,y,z} \right), \\
\tilde P_3^{p,R}\left( {x,y,z} \right) & = \left( {P_3^{p,R} * {P_{3,sys}}} \right)\left( {x,y,z} \right),    
\end{align*}
are the 3D partially-coherent amplitude and phase Fresnel propagators, respectively, and ${P_{3,sys}}\left( {x,y,z} \right)$ is the point-spread function of the imaging system in the reconstructed 3D volume.

Solutions of the 3D PB-CT reconstruction problem can be obtained from \cref{eq3:fbp_ctf_3d_partial_coherent} in exactly the same way as \cref{eq3:beta_fourier_3d,eq3:delta_fourier_3d,eq3:ctf_hom_mu_3d} were obtained from \cref{eq3:fbp_ctf_3d}.

\section{Numerical simulations} \label{s3:numerical_simulation_framework}

\subsection{X-ray CT simulation framework}

The 3D CTF-Hom approximation for phase and amplitude retrieval on post-CT reconstructed images as derived in \cref{s3:derivation_2} will now be referred to as \textit{PostCTFHom}. Similarly, the corresponding 2D method applied to projections prior to CT reconstruction will be referred to as \textit{PreCTFHom}.
In order to evaluate the accuracy and characteristics of the PostCTFHom, a computational simulation platform was developed in the form of a web-based Jupyter Notebook \cite{Kluyver2016JupyterWorkflows}, utilizing the Python \cite{VanRossum2009PythonManual} programming language. Additionally, the \textit{Syris} \cite{Farago2017Syris:Simulation} framework was used and extended upon with the addition of CT reconstruction via the Astra-toolbox \cite{vanAarle2016FastToolbox} package and bespoke TIE and CTF phase retrieval implementations to simulate conventional absorption CT and PB-CT workflows. 

The \textit{Syris} framework and associated packages allowed for the definition of a simple analytic 3D cylindrical model representing a material defined by its complex refractive index at a given X-ray energy. With such a model, the workflow can be used to generate a volume image in addition to contact or PBI projections for a given number of rotation angles with specified resolution and photon statistics. These simulated projections can then be subsequently reconstructed, including phase retrieval as part of the workflow, and quantified with a range of metrics. 

\subsection{Numerical model} \label{s3:single_cylider_numerical_model}

The numerical model defined for these simulations consists of a $1024\;\mu m \times 1024\;\mu m$ rectangle containing a single central $50\;\mu m$ diameter circle composed of a lightweight polymer like \textit{virtual} object in vacuum. To reduce computational complexity, this symmetric 2D model was further simplified to simulating an equivalent set of 1D projections. The virtual materials $\beta$ and $\Delta$ values were calculated using the online database at \href{https://henke.lbl.gov}{henke.lbl.gov} \cite{Henke1993X-Ray1-92}, specifying the required energy, chemical formula and mass density. For the simulations, a fixed X-ray energy, $E = 20\;{\rm{keV}}$ was used throughout. The virtual objects mass density was adjusted to simulate variable X-ray transmission in terms of the associated complex refractive index. For example, a density of $\rho=0.001\;\left(g/cm^3\right)$, corresponds to $\mu=3.647 \times {10^{-2}}\;\left(m ^{{-1}}\right)$, $\beta=1.799 \times {10^{-13}}$, $\Delta=5.922 \times {10^{-10}}$ and maximum phase-shift, $\phi=-0.030 \left(rad\right)$ over the $50\;\mu m$ maximum diameter of the object. The ratio $\gamma = \Delta / \beta$ is 3291.

\subsection{Simulations} \label{s3:simulations}

In this section, we evaluate and validate the PostCTFHom formalism derived above in \cref{s3:derivation_1} with the numerical framework and model defined in the previous sections. We then compare its performance and characteristics with that of conventional absorption CT and the PreCTFHom method. 

\subsubsection{Projection simulation} \label{s3:projection_simulation}

Distributions of transmitted intensity and phase shift were computed analytically using the model definition in \cref{s3:single_cylider_numerical_model} and the following equations for transmitted phase and intensity, respectively.
\begin{align*}
{\varphi _\theta }(x',y) =  - k{P_\theta }\Delta (x',y), \\
{I_\theta }(x',y) = {I_{in}}\exp \left[ { - 2k{P_\theta }\beta (x',y)}\right],
\end{align*}
with $I_{in} = 1$ for these simulations.

The complex amplitude of the transmitted wave was then calculated over a 1D row sampled at a given resolution. In the case of the simulations for this paper, a fine sampling interval of $6.25\times 10^{-2}\;\mu m$ was chosen for the generation of initial projections. Due to the invariance of the model along the rotation \textit{y-axis}, we can employ the simplification of only requiring the generation of a single-row projection, resulting in 1D row-projection arrays of 16384 pixels.

For PBI projections, the Fresnel propagation operator \cite{Paganin2006CoherentOptics} implemented as a 1D Fourier filter was applied to each 1D complex amplitude, with the transfer function, $FG\left( {\xi '} \right) = \exp [-i\pi \lambda R{\xi '^2}]$, with $R$ being the propagation distance. In the case of contact projections, this propagation step was skipped.

In order to simulate finite system resolution, a discrete Gaussian smoothing filter, with a standard deviation ${\sigma^2} = 1\;\mu m$ was applied to the contact or propagated simulated projection.

Optionally, noise may be added to the simulated projection at this point. For this work, a Poisson distribution was chosen to model that of a photon-counting detector. The amount of noise applied is specified by the relative average standard deviation, which is equivalent to ${N_{pp}}^{-\frac{1}{2}}$, with ${N_{pp}}$ being the number of photons registered at the point of average intensity.

Finally, projections were re-sampled, or decimated, with a finite square aperture of, $1\;\mu m$ resulting in a $1024$ pixel 1D row intensity projection. This decimation step also includes an additional application of a Gaussian smoothing filter with ${\sigma ^2}$ equal to the full width at half maximum (FWHM) of $1\;\mu m$, in order to account for a simulated finite detector pixel size.

\begin{figure}[!htbp]
    \centering
        \includegraphics[width=\linewidth]{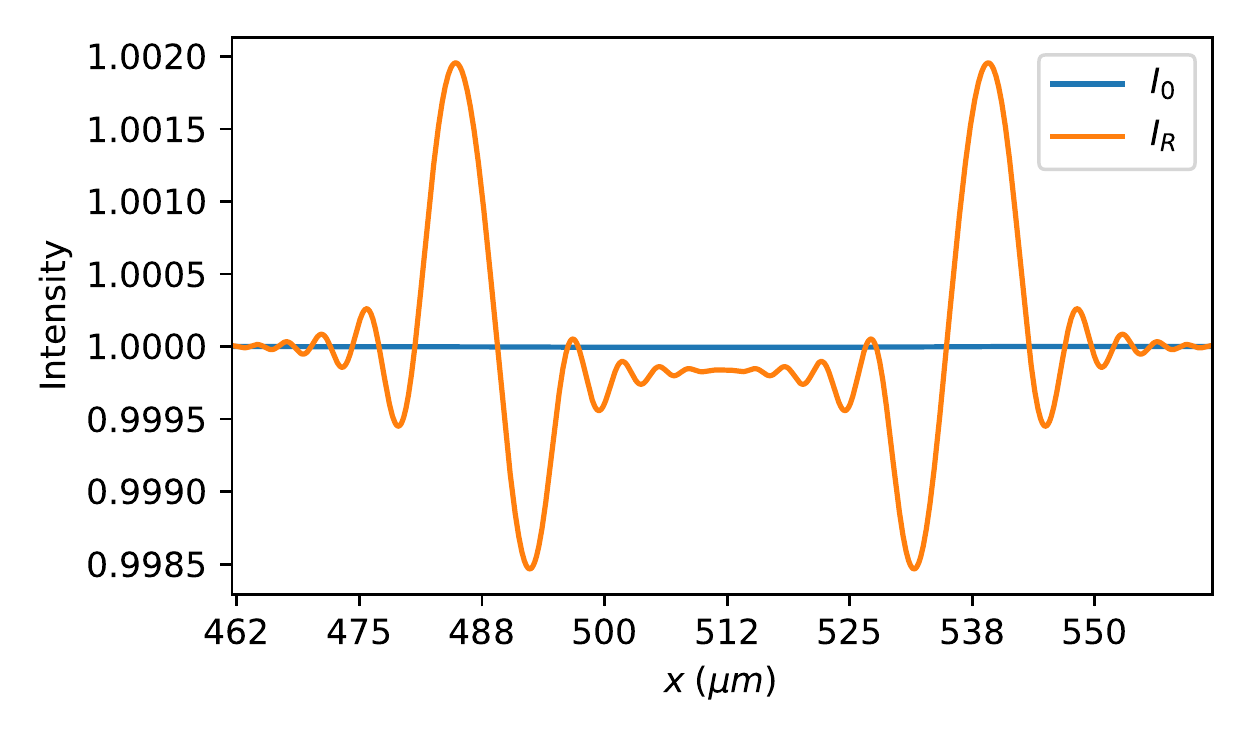}%
        \caption{Simulated projection intensities, $\rho=0.003\;g/cm^3$, $R=0.5\;m$ , $|\phi|= 0.01\;rad$.}
        \label{fig3.simulated_projection_intensity}
\end{figure}

\Cref{fig3.simulated_projection_intensity} shows an example plot of a finely-sampled (pixel size = $6.25 \;\times\;{10^{-2}}\;\mu m$) simulated 1D contact, $I_0$, and propagated, $I_R$ intensity projections of the cylindrical numerical model with  $|\phi|= 0.01\;rad$. $I_R$ has been propagated through a distance of $R=0.5\;m$. The plot of $I_0$ displays practically no detectable contrast, whereas at a propagation distance of $R=0.5\;m$, multiple diffraction fringes are visible in the plot of $I_R$.

\subsubsection{Pre-reconstruction 2D CTF-Hom phase retrieval (PreCTFHom)}\label{s3:pre-ctfhom-2d}

For PreCTFHom, phase retrieval is applied directly to the simulated intensity projections in the form of a Fourier space filter, prior to any subsequent CT reconstruction. As described by Nesterets and Gureyev \cite{Nesterets2014NoiseTomography}, the intensity based form of 2D CTF-Hom phase retrieval is a rearrangement and simplification of \cref{eq3:intensity_r_ctf_weak},
\begin{equation}
    \label{eq3:ctf_hom_2d_intensity}
    {F_2}I_\theta ^0\left( {\xi ',\eta} \right) \cong \frac{I_{in}^{-1}\space{F_2}I_\theta ^R\left( {\xi ',\eta} \right)}{\sgn \left(\gamma\right) \sqrt{\gamma^2 + 1}\sin{\left[A\left( {\xi ',\eta} \right) \right]}},
\end{equation}
where $A\left( {\xi ',\eta} \right) = \pi \lambda R \left(\xi '^2 + \eta^2\right) + \atan{\left(\gamma^{-1}\right)}$. In order to avoid zeroes in the denominator of the sine function in \cref{eq3:ctf_hom_2d_intensity}, the following regularized form \cite{Nesterets2014NoiseTomography} has been adopted,
\begin{equation}
    \label{eq3:ctf_hom_reg_2d_intensity}
    {F_2}I_\theta ^0\left( {\xi ',\eta} \right) \approx \frac{\sgn \left(\gamma\right)I_{in}^{-1} {F_2}I_\theta ^R\left( {\xi ',\eta} \right)\sin{\left[A\left( {\xi ',\eta} \right)\right]}}{\sqrt{\gamma^2 + 1}\left\{ \sin^2{\left[A\left( {\xi ',\eta} \right) \right]} + \epsilon\left({\xi ',\eta}\right) \right\}}.
\end{equation}
where, in the case of positive $\gamma$, the regularization parameter $\epsilon$ is:
\begin{equation}
    \label{eq3:ctf_hom_epsilon}
    \epsilon\left({\xi ',\eta}\right) = \left\{ 
        \begin{array}{l}
            0,\quad A\left( {\xi ',\eta} \right) < \pi/2\\
            \epsilon_0,\quad otherwise.
        \end{array} \right.
\end{equation}
Here, $\epsilon_0 \ll 1$ is a small regularization constant. 

It should be noted, that in the simulations performed in this work, the above 2D expressions were reduced and applied in 1D to the simulated 1D projections of the cylindrical model.

\subsubsection{CT reconstruction} \label{s3:sim_ctf_ct_recon}

To recover the imaginary component of the complex refractive index from the simulated projections, the FBP CT reconstruction algorithm, \cref{eq3:fbp_operator} was applied. Again, due to the symmetrical nature of the cylindrical numerical model, the process is significantly simplified compared with an experimental context. In this case, it is only necessary to reconstruct a single slice of the final 3D volume. As such, a single sinogram was constructed by stacking a set of ${N_{proj}}$ 1D simulated projections where ${N_{proj}}$ was chosen to minimally satisfy the angular Nyquist sampling condition \cite{Hsieh2015ComputedAdvances} where ${N_{proj}} = {N_{pxl}} \pi / 2$, where ${N_{pxl}}$ is the width of the projection in pixels.  

For noise-free scenarios, the sinogram is constructed by simply repeatedly copying the same simulated 1D projection, after phase retrieval is applied for PreCTFHom. In the case of simulated noise, each 1D projection of the ${N_{proj}}$ row sinogram requires the separate application of Poisson noise followed by phase retrieval in the case of PreCTFHom.

A negative logarithm is then applied globally to the fully assembled ${N_{pxl}} \times {N_{proj}}$ pixel sinogram. The final step of FBP CT reconstruction is applied to the sinogram, resulting in  a single ${N_{pxl}} \times {N_{pxl}}$ pixel 2D slice of the reconstructed linear attenuation coefficient $\mu$, which relates to $\beta$ in \cref{eq3:ctf_hom_mu_3d} by the simple multiplication, $\mu=2k \beta $. The fixed $h=1\;\mu m$ pixel size results in ${N_{pxl}} = 1024$ and ${N_{proj}} = 1609$, thus corresponding to an angular step of $180/1609 \approx 0.11^{\circ }$.

To permit a full suite of comparative performance metrics to be calculated, each CT reconstruction phase of the simulation also generates sinograms for unpropagated contact ($I_0$) in addition to propagated projections with and without PreCTFHom applied. Furthermore, noisy and noise-free versions of each are also generated.

\subsubsection{Post-reconstruction 3D CTF-Hom phase retrieval (PostCTFHom)}

As the name implies, PostCTFHom is applied to a sub-volume of the reconstructed 3D volume. As discussed in \cite{Thompson2019FastTomography}, successful application of 3D post CT reconstruction phase retrieval such as PostTIEHom and PostCTFHom is reliant on a considered choice of the 3D region of interest (ROI), ${\Omega _m}$. Firstly, ${\Omega _m}$ should fully contain the object under investigation. Secondly, ${\Omega _m}$ should be chosen to consider the width of the 3D point-spread function (PSF) of CTF-Hom. For the simulations performed in this work, several sizes were considered, the full reconstructed volume, ${\Omega _{1024}}$, and sub-volumes ${\Omega _{512}}$, ${\Omega _{256}}$, and ${\Omega _{128}}$, which just enclosed the reconstructed object.
Once again, due to the inherent symmetries of the simulated object, it was possible to simplify the computation of PostCTFHom to that of the application of the 2D CTF-Hom Fourier filters described in \cref{s3:pre-ctfhom-2d} by \cref{eq3:ctf_hom_2d_intensity,eq3:ctf_hom_reg_2d_intensity} (for the regularized version) to the reconstructed 2D slice.

\subsubsection{Evaluation metric} \label{s3:ctf_evaluation_metrics}

To quantify and compare the performance of the evaluated phase retrieval methods, the root mean squared error (RMSE) was used, which is defined between a target, $x$ and reference, $x'$ image, by:

\begin{equation}
    \label{eq3:rmse}
    RMSE = \sqrt{\frac{1}{N}{\sum_{i}^{N}{\left(x_i'-x_i\right)^2}}}.
\end{equation}

\subsection{Results}\label{s3:results}

The fixed simulation parameters, described in \cref{s3:simulations} outlining the simulation framework have been specifically selected to be consistent with an experimental micro-PB-CT imaging scenario.  

In addition to the fixed simulation parameters,  energy ($20\;{\rm{keV}})$, sample size $(50\;\mu m)$, material and pixel size $(1\;\mu m)$, a range of variable simulation parameters can be passed to the simulation framework. The number of potential degrees of freedom within this parameter space presents a challenge for rigorous analysis. Given this, for the scope of the analysis and results presented in this research, the parameter set has been constrained to a single specific simulated imaging scenario, namely with $R=1m$, $|\phi|=0.01\;rad$, $1.0\%$ noise ($N_{pp} = 10000$) and $\epsilon=0.1$. The choice of $\epsilon$ is examined in more detail in \cref{s3:PostCTFHom}. Finally, for the post-reconstruction region of interest (ROI) size, ${\Omega _m}$, a range of values are used and are discussed in \cref{s3:ctf_roi}.

All phases described in \cref{s3:simulations} correspond to a single simulation run for the given set of fixed and variable parameters, producing a set of images and performance metrics as specified in \cref{s3:ctf_evaluation_metrics}, allowing a direct comparison between the PreCTFHom and PostCTFHom methods.

\subsubsection{Pre-CT reconstruction phase retrieval}\label{s3:PreCTFHom}

\begin{figure}[!htbp]
    \centering
        \includegraphics[width=\linewidth]{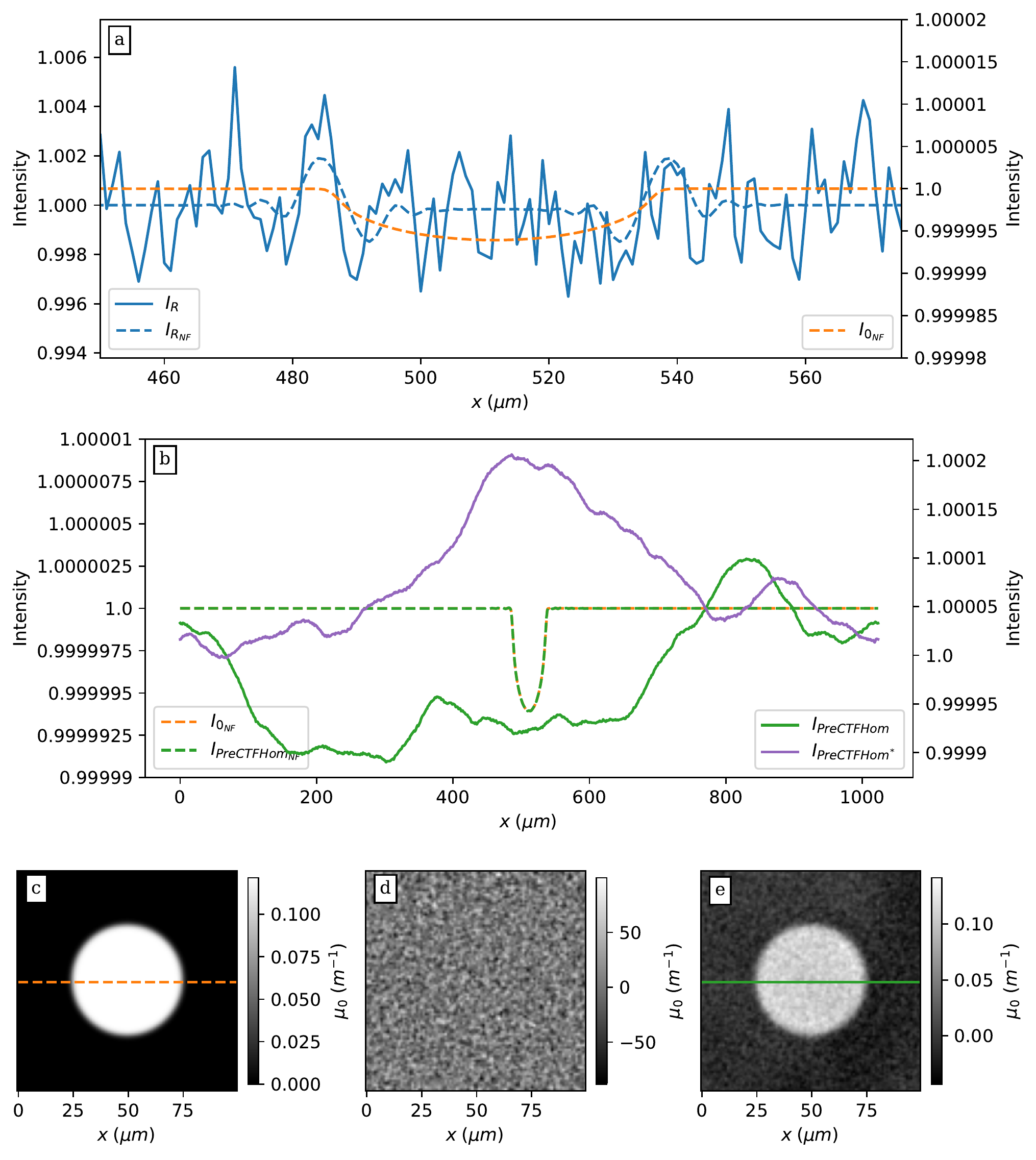}
        \caption{ Plot a) Intensity profiles of, $I_{{0_{{NF}}}}$, $I_{{R_{{NF}}}}$ and $I_R$. b) Intensity profiles of, $I_{{0_{{NF}}}}$, $I_{PreCTFHom_{NF}}$,$I_{PreCTFHom}$ and $I_{PreCTFHom^{*}}$. Images of CT reconstructed slices, c)  $\mu_{0_{NF}}$, d) $\mu_{0}$, e) $\mu_{PreCTFHom}$. See \cref{s3:PreCTFHom} for details.}
        \label{fig:combined_PreCTFHom}
\end{figure}

\Cref{fig:combined_PreCTFHom} presents a series of plots and images corresponding to simulations of the PreCTFHom method. \Cref{fig:combined_PreCTFHom}a shows reference noise-free intensity profiles of, contact $I_{{0_{{NF}}}}$ and propagated, $I_{{R_{{NF}}}}$ projections. For this configuration with $R=1m$ and $|\phi|=0.01\;rad$, multiple diffraction fringes and enhanced contrast are clearly visible in the plot of $I_{{R_{{NF}}}}$ when compared to the intensity/contrast of $I_{{0_{{NF}}}}$.  In contrast, the plot of $I_R$, displays no discernible object structure in the presence of $1.0\%$ noise.  

\Cref{fig:combined_PreCTFHom}b illustrates the result of applying CTF-Hom phase retrieval (PreCTFHom) to the simulated propagated projections shown previously in \cref{fig:combined_PreCTFHom}a.  $I_{PreCTFHom}$ and $I_{PreCTFHom^{*}}$ illustrates two instances of the application of CTF-Hom phase retrieval applied to two propagated projections, $I_R$ with different noise profiles. Interestingly, at this noise level $(1\%)$, CTF-Hom phase retrieval is not able to recover any recognizable structure of the object in either example. However, it is evident that there has been a change in the statistical properties of the noise, with a shift from uncorrelated to correlated noise. For reference, plots of $I_{PreCTFHom_{NF}}$ with PreCTFHom phase retrieval applied to $I_{R_{NF}}$ and $I_{0_{NF}}$ are displayed. As expected, in noise-free conditions, the former displays an almost perfect reconstruction, thus overlapping the plot of  $I_{0_{NF}}$.

\Cref{fig:combined_PreCTFHom}(c-e) present a series of CT-reconstructed slices computed via sinograms assembled from projections with intensity profiles of $I_{{0_{{NF}}}}$ and $I_{PreCTFHom_{NF}}$ as shown in \cref{fig:combined_PreCTFHom}a and \cref{fig:combined_PreCTFHom}b. The coloured horizontal line on the slice images indicates the displayed profile. Note, a corresponding plot for $I_{{0}}$ is not shown.
\Cref{fig:combined_PreCTFHom}c illustrates a CT reconstruction from noise-free contact projections, $\mu_{0_{NF}}$ providing a reference image of the simulated object. At the other end of the spectrum, \cref{fig:combined_PreCTFHom}d shows the result of reconstructing a slice using noisy contact projections, $\mu_{0}$, and highlighting the inability of conventional absorption CT to recover the weakly attenuated object from projections with this level of noise.
\Cref{fig:combined_PreCTFHom}e displays a CT-reconstructed slice from projections with PreCTFHom phase retrieval applied, $\mu_{PreCTFHom}$ demonstrating the visually successful recovery of the object. Importantly, CTF-Hom, despite some residual noise both within and outside the object, recovers sharp edges between the object and background without significant artefacts introduced. The corresponding profile for this reconstruction is shown in \cref{fig:combined_PostCTFHom}e. It is quite remarkable, when visually comparing the images of \cref{fig:combined_PreCTFHom}d and \cref{fig:combined_PreCTFHom}e that projection-based phase retrieval followed by FBP CT reconstruction is able to reconstruct the object to such a degree when one considers the lack of obvious structure in projections even after phase retrieval. Specifically, in the context of this simulated scenario, much of this recovered detail can be attributed to the large number of projections $(\approx 1600)$ used for CT reconstruction step, where the now correlated noise in projections cancels out over the full projection space.

\subsubsection{Post-CT reconstruction phase retrieval}\label{s3:PostCTFHom}

\begin{figure}[!htbp]
    \centering
        \includegraphics[width=\linewidth]{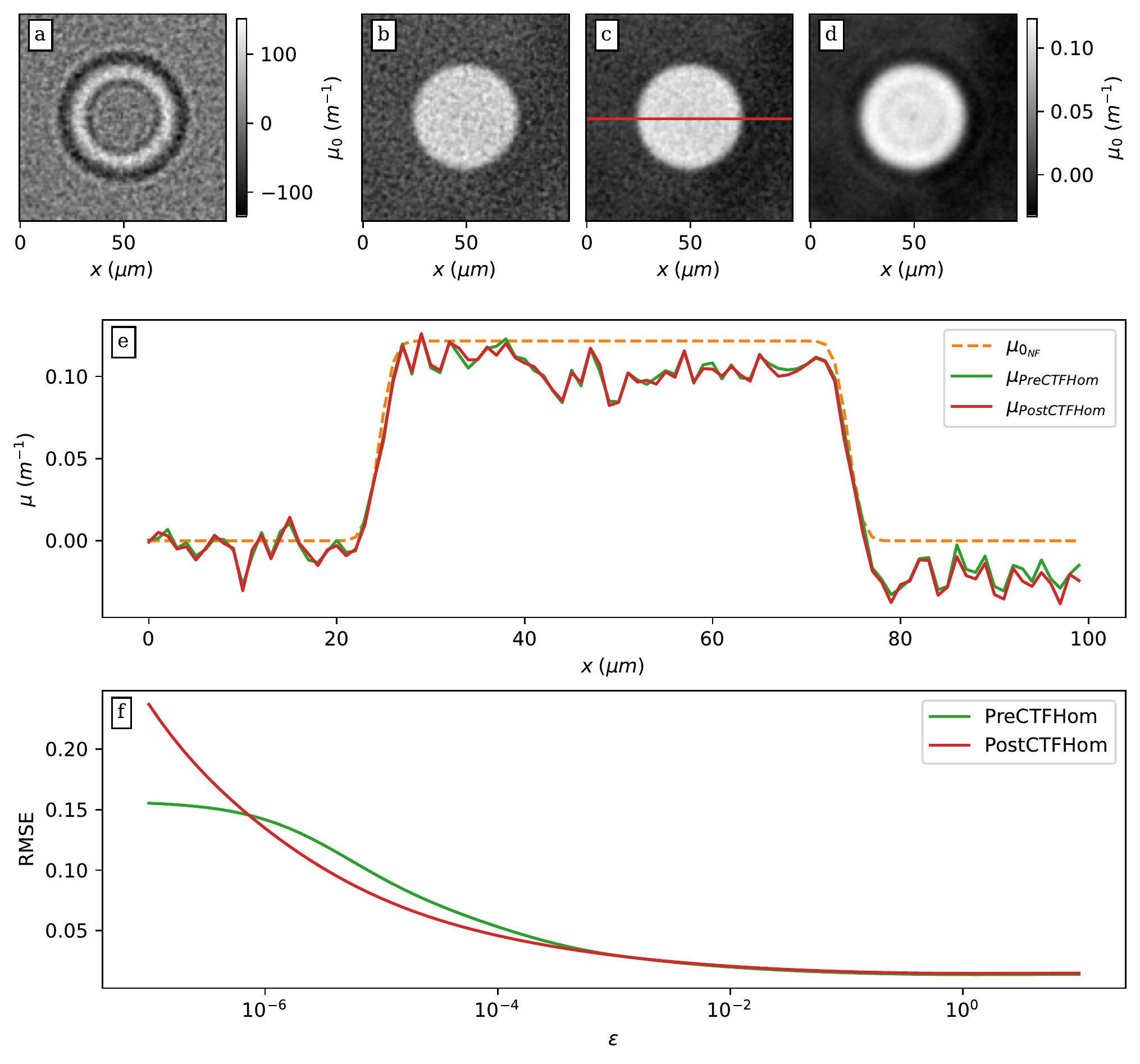}
        \caption{Images of CT reconstructed slices,  a) $\mu_R$, b) $\mu_{PostCTFHom}, \epsilon=0.01$, c) $\mu_{PostCTFHom}, \epsilon=0.1$, d) $\mu_{PostCTFHom}, \epsilon=1.0$. e) CT reconstruction profiles, $\mu_{0_{NF}}$, $\mu_{PreCTFHom}$ and $\mu_{PostCTFHom}$, $\epsilon=0.1$.  f) Reconstruction error (RMSE),  $\mu_{PreCTFHom}$ and $\mu_{PostCTFHom}$ vs $\epsilon$.}
        \label{fig:combined_PostCTFHom}
\end{figure}

\Cref{fig:combined_PostCTFHom} presents results from the new PostCTFHom method. For this set of simulations,  phase retrieval using an ROI of the full reconstructed slice, $1024 \times 1024$ pixels was used. This permits a direct comparison against the full projection phase retrieval PreCTFHom method. Variation of the PostCTFHom ROI size is examined in \cref{s3:ctf_roi}.

The top row of \cref{fig:combined_PostCTFHom}, images a-d displays a set of reconstructed CT slices. \cref{fig:combined_PostCTFHom}a shows the reconstruction of noisy propagated projections, $\mu_R$. Unlike its corresponding noisy projection $I_R$, without phase retrieval, as seen in the profile plot (\cref{fig:combined_PreCTFHom}a), CT reconstruction manages to recover multiple diffraction fringes which are subsequently exploited by PostCTFHom phase retrieval. A regularized form of \cref{eq3:ctf_hom_mu_3d}, with the same regularization used in \cref{eq3:ctf_hom_reg_2d_intensity}, was implemented in the PostCTFHom method. Therefore, this method also requires the selection of an appropriate regularization parameter, $\epsilon$. For the simulations shown previously in \cref{s3:PreCTFHom} a fixed $\epsilon=0.1$ was chosen for application with the PreCTFHom method. The CT reconstructed slices presented in \cref{fig:combined_PostCTFHom}b, c and d demonstrate the result of using different values of $\epsilon$, namely $\epsilon=$ 0.01, 0.1 and 1.0 respectively. \cref{fig:combined_PostCTFHom}e shows a profile plot of the central row of the CT reconstructed slice of \cref{fig:combined_PostCTFHom}c, as indicated by the overlaid red horizontal line on \cref{fig:combined_PostCTFHom}c, $\mu_{PostCTFHom}$ with $\epsilon=0.1$. Additionally, \cref{fig:combined_PostCTFHom}e plots profiles of the associated $\mu_{PreCTFHom}$ result and $\mu_{0_{NF}}$ for reference. Finally, \cref{fig:combined_PostCTFHom}f plots the CT reconstruction error (RMSE) of $\mu_{PreCTFHom}$ and $\mu_{PostCTFHom}$ against $\mu_{0_{NF}}$ over eight orders of magnitude of $\epsilon$, with $1 \times 10^{-7}<=\epsilon<=10$. The plot of \cref{fig:combined_PostCTFHom}f demonstrates the influence of $\epsilon$ with CTF-Hom phase retrieval and its effect on reconstruction quality. Aside from a  divergence at very small $\epsilon$ values,  $\epsilon<1\times 10^{-6}$, there is little appreciable difference between the two methods for $\epsilon=0.1$. PostCTFHom displays a marginally lower overall RMSE for $1 \times 10^{-6}<\epsilon<1 \times 10^{-3}$. Visually, for CT reconstructed slices, this difference is imperceptible. 
The primary benefit of increasing $\epsilon$ for CTF-Hom, which is clearly visible in the reconstructed slices of \cref{fig:combined_PostCTFHom}b, c and d is in the suppression of noise. In the context of the given simulation parameters, $\epsilon=0.1$, corresponding to \cref{fig:combined_PostCTFHom}c, represents a good compromise of noise suppression and retention of feature integrity. \cref{fig:combined_PostCTFHom}b, where $\epsilon=0.01$ is relatively too noisy. \cref{fig:combined_PostCTFHom}d, with $\epsilon=1.0$ has suppressed noise significantly compared to the other two, however the feature itself now appears over smoothed and with the introduction of faint fringes around the centre and beyond the edge, this value of $\epsilon$ also corresponds to the flat region of the plot in \cref{fig:combined_PostCTFHom}f where the large regularization value of $\epsilon$ results in CTF-Hom closely resembling TIE-Hom, and its inherent low-pass filter properties.

\subsubsection{Effect of PostCTFHom region of interest}\label{s3:ctf_roi}

\begin{figure}[!htbp]
    \centering
        \includegraphics[width=\linewidth]{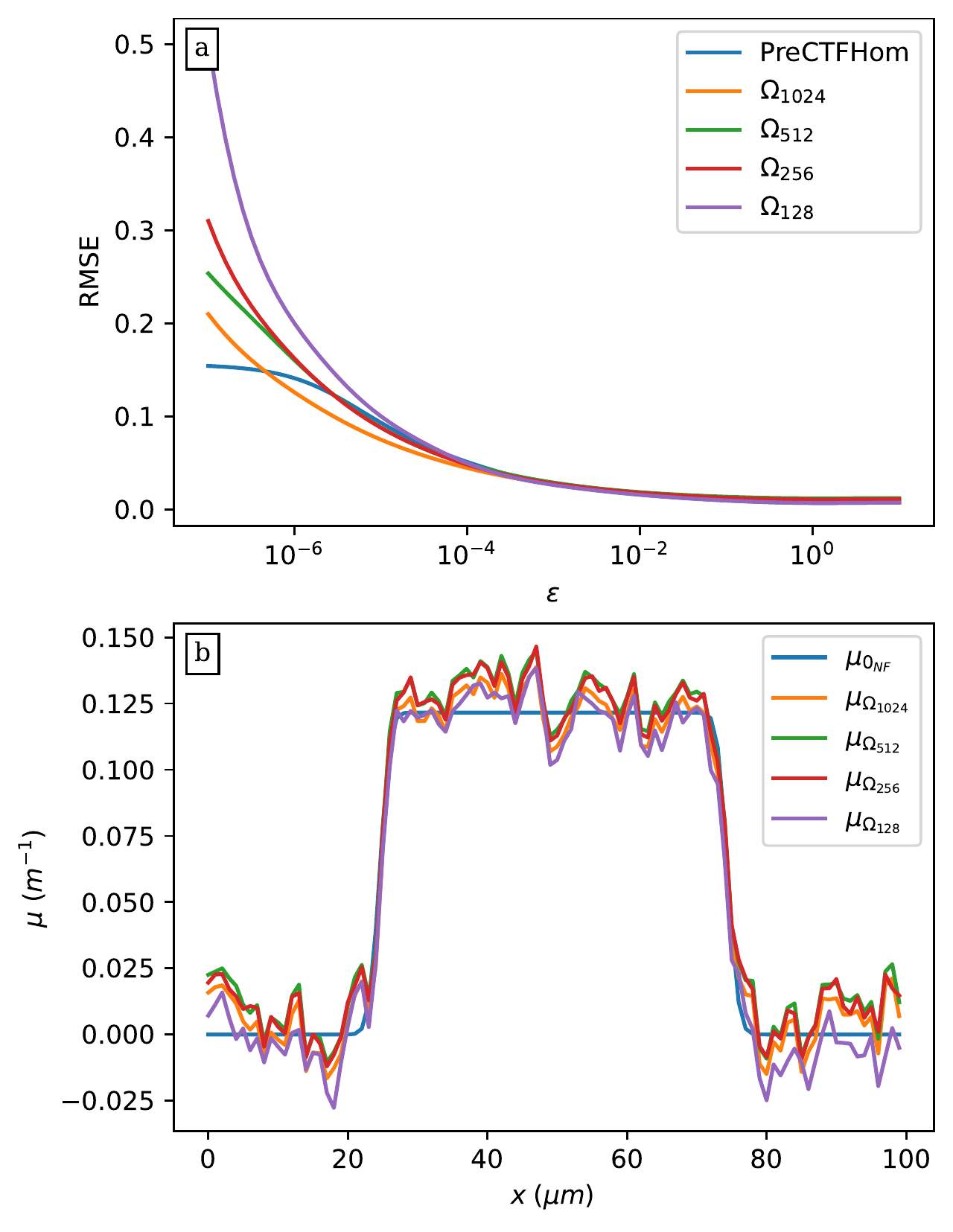}
        \caption{Comparison of varying PostCTFHom ROI widths of 1024, 512, 256 \& 128 pixels. Plot a) PostCTFHom RMSE vs $\epsilon$. b) PostCTFHom reconstruction profiles of varying ${\Omega _m}$}
        \label{fig:PostCTFHom_roi}
\end{figure}

\Cref{fig:PostCTFHom_roi} presents the results of varying the PostCTFHom ROI size, ${\Omega _m}$ in conjunction with the CTF-Hom regularization parameter, $\epsilon$. A range of square regions for ${\Omega _m}$ were chosen, corresponding to widths of 1024 (full width), 512, 256 and 128 pixels respectively. \Cref{fig:PostCTFHom_roi}a plots the reconstruction error (RMSE) between $\mu_{0_{NF}}$ and the corresponding PostCTFHom reconstruction at the four given values of ${\Omega _m}$. An equivalent plot of $\mu_{PreCTFHom}$ is provided for comparison. Evident from the plot is the relative insensitivity of the size of the chosen value of ${\Omega _m}$ to the overall image quality of PostCTFHom phase retrieval, even for a relatively small 128 pixel width region which extends only a small distance beyond the boundary of the reconstructed object itself. Similar to the result shown in \cref{fig:combined_PostCTFHom}f, again, variation in RMSE for all values of ${\Omega _m}$ is observable only for small values of $\epsilon\le1\times 10^{-6}$. Above this, the calculated RMSE is virtually identical between the PreCTFHom and PostCTFHom for all ${\Omega _m}$.

\Cref{fig:PostCTFHom_roi}b plots PostCTFHom reconstruction profiles of the central row of slices for the range of ${\Omega _m}$ values with $\epsilon=0.1$. A plot of $\mu_{0_{NF}}$ is also displayed for comparison. As implied by the results shown in \cref{fig:PostCTFHom_roi}a, the reconstruction profiles are almost identical for all ${\Omega _m}$.

It should be noted that when applying PostCTFHom to a localised ROI, it is possible to induce a small constant shift in the reconstructed $\mu$ value due to a potential different background DC component within the selected ROI compared to the full field of view. In order to compensate for this phenomenon, the background mean value was calculated in an area outside the object with the ROI and subtracted from the resultant phase retrieved image.

The ability to apply PostCTHom to a localised CT reconstructed ROI is also the source of potentially significantly reduced computation costs compared to PreCTFHom and other projection-based phase retrieval methods\cite{Thompson2019FastTomography}. Essentially, for PostCTHom the phase retrieval step is computed via a single 3D Fourier filter operation on a localised and variably sized sub-image ${\Omega _m}$. By comparison, PreCTFHom requires the fixed computation of $N_{proj} \times {N_{pxl}}^2$ pixel 2D Fourier filter operations. As can be seen, when the linear size of the ROI is less than $N_{pxl}$ the computational cost is reduced proportionally.

\section{Conclusion}

In this paper, we have developed a method for PB-CT reconstruction of the 3D distribution of the complex refractive index in weakly scattering objects from multiple 2D transmission images collected in the Fresnel region (at some free-space propagation distance from the object) using coherent or partially-coherent incident X-ray beams at different illumination directions. It is instructive to summarize the assumptions that led to the main results of this paper,  \cref{eq3:ctf_contrast,eq3:fbp_ctf_3d,eq3:contrast_ctf_approx,eq3:fbp_ctf_3d_partial_coherent}.

In order to derive \cref{eq3:ctf_contrast,eq3:fbp_ctf_3d} the following assumptions were made.

\begin{enumerate}
    \item A fully-coherent incident plane wave, ${U_{in}}({\bf{r}}) = I_{in}^{1/2}\exp (ikz)$, was assumed
    \item The projection approximation was applied for the interaction of the incident wave with the object, \cref{eq3:complex_scalar_transmission,eq3:projection_operator}.
    \item Paraxial approximation (Fresnel diffraction) was assumed for the free-space propagation between the object and the detector, \cref{eq3:paraxial_wave}.
    \item  Weak absorption in the object and Guigay’s condition (slow variation) for the phase, \cref{eq3:weak_attenuation_phase}, was assumed.
\end{enumerate}

In order to derive \cref{eq3:contrast_ctf_approx,eq3:fbp_ctf_3d_partial_coherent}, the assumptions 1 and 4 have been relaxed and replaced with the following assumptions.

\begin{enumerate} \label{s3:conclusion}
    \setcounter{enumi}{4}
    \item  Partially-coherent illumination of the object was assumed instead of the fully-coherent illumination.
    \item In addition to the weak component, as in item 4 above, the attenuation function of the object was allowed to have a strong but slowly-varying component, \cref{eq3:attenuation_sm_sl}. This latter component of the attenuation function was assumed to be slowly varying on the length scale of the width of the 2D partially-coherent Fresnel propagator, $\tilde P_2^R\left( {x',y} \right) = \left( {P_2^R * {P_{sys}}} \right)\left( {x',y} \right)$.
\end{enumerate}

The PB-CT reconstruction formulae, \cref{eq3:beta_fourier_3d,eq3:delta_fourier_3d}, have been derived under the above assumptions 1-4 in the case of a generic weakly-scattering object and two images per illumination direction collected at different object-to-detector distances and a range of illumination directions. We also gave a corresponding solution, \cref{eq3:ctf_hom_mu_3d}, for the 3D reconstruction of monomorphous objects from a single image per illumination direction. Similar equations can be easily derived from \cref{eq3:contrast_ctf_approx} in the case of partially-coherent illumination and strongly absorbing samples satisfying the assumptions 5-6.

The simulated results in \cref{s3:numerical_simulation_framework}, demonstrate that the PostCTFHom method provides a fast and stable phase retrieval method, producing results consistent with that of the pre-reconstruction projection-based variant. Moreover, PostCTFHom, being an extension (for greater propagation distances) to the PostTIEHom method \cite{Thompson2019FastTomography}, can be applied to a localised sub-volume of the full CT-reconstructed volume. Results presented in \cref{s3:ctf_roi} indicate that the choice of dimensions for ${\Omega _m}$, can be highly localised to just beyond the spatial boundary of the object of interest without a significant loss of quality whilst reducing the time and complexity of computation. As also discussed in \cite{Thompson2019FastTomography}, this localised phase retrieval permits a material specific $\gamma$ local value to be applied and tuned for optimal results. The PostCTFHom method introduces the additional CTF-Hom regularization parameter, $\epsilon$ which can also be independently tuned to maximise the reconstruction quality of localised regions.  Computationally, PostCTFHom implements a different 3D Fourier filter method from PostTIEHom but exhibits the same computational performance characteristics as described in  \cref{s3:ctf_roi} and the \textit{Computational analysis} section of \cite{Thompson2019FastTomography}, whereby computational performance depends on the size of the region of interest chosen. 





\end{document}